\documentclass[10pt,twocolumn,amsmath,amssymb,amsfonts,aps,prd,superscriptaddress,showpacs,nofootinbib]{revtex4-2}
\usepackage{booktabs, graphicx, hyperref, newtxtext, newtxmath}
\usepackage[T1]{fontenc}
\usepackage[dvipsnames]{xcolor}

\DeclareUnicodeCharacter{2212}{-}

\hypersetup{colorlinks=true, linkcolor=ForestGreen, citecolor=ForestGreen, filecolor=BurntOrange, urlcolor=Maroon}

\newcommand{\amc}{{\sc MadGraph5}\_a{\sc MC@NLO}}
\newcommand{\fr}{{\sc Feyn\-Rules}}

\newcommand{\mspin}{{\sc MadSpin}}
\newcommand{\mw}{{\sc MadWidth}}
\newcommand{\ma}{{\sc MadAnalysis~5}}
\newcommand{\py}{{\sc Pythia~8}}
\newcommand{\fj}{{\sc FastJet}}
\newcommand{\sfs}{{\sc SFS}}
\newcommand{\pyhf}{{\sc PyHF}}
\def\ysS{y_{1S}}
\def\yoS{y_{8S}}
\def\sing{S_1}
\def\oct{S_8}
\newcommand{\scr}[1]{\ensuremath{\mathcal{#1}}}

\newcommand*{\CPthree}{~\small \textit{Centre  for  Cosmology,  Particle  Physics  and  Phenomenology  (CP3), Université  catholique  de  Louvain,  1348  Louvain-la-Neuve,  Belgium}}
\newcommand*{\UCBL}{~\small \textit{Universit\'e Claude Bernard Lyon 1, CNRS/IN2P3,
Institut de Physique des 2 Infinis de Lyon, UMR 5822, F-69622, Villeurbanne, France}}
\newcommand*{\LPTHE}{~\small \textit{Laboratoire de Physique Th\'{e}orique et Hautes \'{E}nergies (LPTHE), UMR 7589, Sorbonne Universit\'{e} \& CNRS, 4 place Jussieu, 75252 Paris Cedex 05, France}}

\begin{document}

\title{Boosting Beyond: A Novel Approach to Probing Top-Philic Resonances at the LHC}

\author{\vspace{0.5cm} Luc Darmé}
\affiliation{\UCBL}
\author{\vspace{0.5cm} Benjamin Fuks}
\affiliation{\LPTHE}
\author{Hao-Lin Li}
\affiliation{\CPthree}
\author{Matteo Maltoni}
\affiliation{\CPthree}
\author{Olivier Mattelaer}
\affiliation{\CPthree}
\author{Julien Touchèque}
\affiliation{\CPthree}

\begin{abstract}
  We introduce a novel search strategy for heavy top-philic resonances that induce new contributions to four-top production at the LHC. We capitalize on recent advances in top-tagging performance to demonstrate that the final state, that is expected to be boosted based on current limits, can be fully reconstructed and exploited. Notably, our approach promises bounds on new physics cross-sections that are a few to 60 times stronger than those obtained with existing searches, showcasing its unprecedented effectiveness in probing top-philic new physics.
\end{abstract}

\maketitle

\textbf{\textit{Introduction}} -- With the third run of the LHC, both the ATLAS and CMS experiments are poised to make significant progress in the detection of heavy top-philic resonances with four-top probes. These resonances, prevalent in various new physics scenarios (see, \textit{e.g.}, \cite{Lillie:2007hd, Pomarol:2008bh, Choi:2008ub, Kumar:2009vs, Dev:2014yca, Cacciapaglia:2015eqa, Cheng:2016tlc, Fox:2018ldq, Carpenter:2020evo, Cacciapaglia:2024wdn}), typically lead to an increased rate of four-top events through their QCD-induced pair production and associated production with a $t\bar{t}$ pair.

Previous LHC analyses focusing on four-top production~\cite{CMS:2017ocm, ATLAS:2018alq, CMS:2019rvj, ATLAS:2023taw} have yielded valuable insights, particularly enabling investigations of the top Yukawa coupling and serving to constrain new top-philic states. Building on this, our prior works~\cite{Fuks:2012im, Beck:2015cga, Calvet:2012rk, Darme:2018dvz, Darme:2021gtt} have demonstrated that slight modifications to existing four-top searches can provide a potent tool for probing top-philic new physics. Specifically, we have shown that new physics contributions to the four-top production cross section at 13~TeV are constrained to approximately 15~fb. Additionally, we anticipate modest improvements in sensitivity when scaling our predictions for the high-luminosity phase of the LHC (HL-LHC) due to the large background uncertainties, that arise when the four-top final state is not fully reconstructed.

In light of these constraints, it becomes evident that substantial improvements in sensitivity could only be achieved through a dedicated approach focusing on boosted top quarks. The present study pursues a dual objective within this context. First, we establish the feasibility of fully reconstructing a final state featuring four boosted top quarks with high efficiency, thereby enabling the potential observation of the corresponding new physics signal amidst the overwhelming Standard Model (SM) background. Secondly, we devise a novel search strategy capitalizing on recent advances in top-tagging techniques~\cite{Gerbush:2007fe, ATLAS:2015ddu, ATLAS:2022qby, CMS:2017wyc}, aimed at improving bounds on top-philic new physics. Utilizing two illustrative scenarios with top-philic scalars, we showcase that our strategy can substantially strengthen cross-section limits, improving by a factor of a few for color-singlet states and up to 60 for color-octet particles compared to existing methods, reaching sub-fb sensitivity.

\smallskip

\textbf{\textit{Theoretical framework and analysis toolchain}} -- To showcase the potential of our innovative search strategy for top-philic particles at the LHC in the four-top final state, we examine two categories of simplified models. We choose them to be invariant under the $SU(3)_c \times U(1)_\mathrm{em}$ gauge group, referring to~\cite{Darme:2021gtt} for a contextualization within the full electroweak symmetry group. In the first scenario, we augment the SM field content with a color-singlet scalar $\sing$, while in the second scenario, we introduce a color-octet scalar $\oct$. The new physics contributions to the Lagrangians of these respective models are~\cite{Darme:2021gtt}:
\begin{align}\label{eq:Lss}
  \!\!\scr{L}_{\sing} &= \frac{1}{2} \partial_\mu \sing \partial^\mu \sing- \frac{1}{2} m_{\sing}^2 \sing^2 +  \ysS\bar{t}   \,\sing ~ \! t\,,\\
  \!\!\scr{L}_{\oct} &= \frac{1}{2} D_\mu \oct^A D^\mu \oct^A -  \frac{1}{2} m_{\oct}^2 \oct^A \oct^A  + \yoS \bar{t}  \, T^A \oct^A ~ \! t \,.
\end{align}
The first two terms in these Lagrangians represent gauge-invariant kinetic and mass terms for the new fields, while the last term describes their interaction with the SM top quark via coupling parameters $\ysS$ and $\yoS$. Additionally, the explicit indices $A$ denote (summed) $SU(3)_c$ adjoint indices. For simplicity, we focus solely on scalar interactions, as the $CP$ nature of the new interactions is not anticipated to impact our results. 

The main phenomenological difference between these two models lies in the possibility for strong production of a pair of color-octet states ($pp \!\rightarrow\! \oct \oct$), while the primary production channel for the color-singlet solely involves the ``associated'' production of the new state with a $t\bar{t}$ pair ($pp \!\rightarrow\! t \bar{t} \sing$), a production mode present in both model classes. Predictions related to the former process remain independent of $\yoS$, resulting in stringent model-independent constraints on moderately heavy color-octet states. This stands in contrast to the latter process, where the production rate can be suppressed by decreasing the new physics couplings. Nevertheless, in both scenarios the primary signature of the model in light of current bounds consists of LHC events featuring four boosted top quarks. Identifying and characterizing these events is the focus of the present study. Depending on the color representation of the scalar, these simplified models can arise from various UV-complete setups, for instance two-Higgs doublet models~\cite{Branco:2011iw,Dev:2014yca,Kraml:2019sis}, non-minimal supersymmetric models~\cite{Salam:1974xa,Fayet:1974pd,Fayet:1975yi,AlvarezGaume:1996mv,Fox:2002bu,Plehn:2008ae,Choi:2008ub,GoncalvesNetto:2012nt,Fuks:2012im,Calvet:2012rk,Benakli:2014cia,Beck:2015cga,Kotlarski:2016zhv,Darme:2018dvz,Carpenter:2020hyz,Carpenter:2020evo} or in models featuring compositeness~\cite{Lillie:2007hd,Pomarol:2008bh,Zhou:2012dz,Cacciapaglia:2015eqa,Liu:2019bua,Cacciapaglia:2020vyf}. Large top-quark couplings typically arise in the latter cases, while supersymmetric scenarios have smaller loop-induced couplings.

To achieve this, both models were implemented into \fr~\cite{Alloul:2013bka} to generate a UFO library~\cite{Darme:2023jdn} suitable for calculations at leading order (LO). Hard-scattering events were generated using \amc~\cite{Alwall:2014hca}, incorporating inclusive decays of unstable final-state particles via \mspin~\cite{Artoisenet:2012st} and \mw~\cite{Alwall:2014bza}, and then matched with parton showers and hadronization modeled by \py~\cite{Bierlich:2022pfr}. Additionally, we simulated the response of a typical LHC detector using the \sfs\ framework~\cite{Araz:2020lnp} of \ma~\cite{Conte:2018vmg}, relying on the default ATLAS parameterization with updated performance~\cite{ATLAS:2019qmc, ATLAS:2020auj, ATLAS:2016gsw}. The use of \ma\ further facilitated event reconstruction using \fj~\cite{Cacciari:2011ma}, and the implementation of the preliminary analysis cuts.

Signal events comprise four boosted top quarks, and their observation necessitates control over various sources of SM background. The primary contributions arise from the production of a top-antitop pair in association with additional bosons and/or jets. Specifically, the dominant background component consists of the production of a top-antitop pair in association with two extra jets ($pp\!\to\! t \bar t j j$), each jet mimicking a boosted top quark. Additionally, the processes $p p \!\to\! t \bar{t} V$ and $p p \!\to\! t \bar{t} V j$ (with $V\!=\!W$, $Z$) contribute, albeit to a lesser extent.\footnote{We separately consider the $t \bar{t} V$ and $t \bar{t} Vj$ background contributions. This follows the analysis requirements, which dictate jets mistagged as tops to be very hard. Consequently, the generated samples do not overlap.} The cross sections for the production of four tops ($p p \!\to\! t \bar t t \bar{t}$) and three tops ($pp\!\to\! t\bar{t}t/t\bar{t}\bar{t}$) are at least two orders of magnitude smaller. Despite their expected lower significance, we include these backgrounds in our analysis. Conversely, we have verified that the multijet and $t\bar{t} VV$ backgrounds are negligible after the selection cuts and top tagging procedure described below. Table~\ref{tab:Xsec} presents cross sections for the most relevant background contributions, both at LO and next-to-leading order (NLO) in QCD, at a center-of-mass energy $\sqrt{S}=$ 13 TeV (a value used throughout this work). Background simulations are conducted with the toolchain introduced above. 

\begin{table}\renewcommand{\arraystretch}{1.4}\setlength{\tabcolsep}{3pt}
\begin{tabular}{c|ccc|ccc}
Process & $\sigma$(LO) & Scale & PDF & $\sigma$(NLO) & Scale & PDF \\ 
\toprule
$t \bar{t} j j $ & 354 &  $^{+ 62\%}_{- 35\%}$ & $\pm~ 5.8\%$ &
   352 & $^{+ 3.7\%}_{- 13\%}$ & $\pm~ 2.6\%$ \\   
   
$t \bar{t} W $ & 0.376 & $^{+23\%}_{-17\%}$ & $\pm~3.9\%$ & 
   0.565 & $^{+ 8.3\%}_{- 8.3\%}$ & $\pm~ 1.8\%$ \\
   
$t \bar{t} W j $ & 0.329 & $^{+39\%}_{-26\%}$ & $\pm~2.1\%$ & 
   0.452  & $^{+ 8.1\%}_{- 12\%}$ & $\pm~ 1.2\%$ \\
   
$t \bar{t} Z $ & 0.563 &  $^{+ 31\%}_{- 22\%}$ & $\pm~4.8\%$ &
   0.756 & $^{+ 9.2\%}_{- 11\%}$ & $\pm~ 2.1\%$ \\
   
$t \bar{t} Z j $ & 0.639 & $^{+ 47\%}_{- 30\%}$ & $\pm~ 6.5\%$ & 
    0.672 & $^{+ 2.6\%}_{- 9\%}$ & $\pm~ 2.5\%$ \\
   
$t \bar{t} t \bar{t}$ & 0.00612 & $^{+ 65\%}_{- 37\%}$ & $\pm~ 13\%$ & 
   0.00920 & $^{+ 28\%}_{- 24\%}$ & $\pm~ 6.0\%$ \\
   
$t \bar{t} t + t\bar{t}\bar{t}$ & 0.00155  & $^{+ 22\%}_{-17\%}$ & $\pm~ 13\%$ &
   0.00201 &  $^{+20 \%}_{-19 \%}$ &  $\pm~7.5 \%$ \\
\end{tabular}
\caption{LO and NLO cross sections, in pb, for the dominant SM background contributions relevant to our analysis, given at LO and NLO in QCD and for a center-of-mass energy of 13 TeV. Predictions rely on the NNPDF2.3NLO set of parton densities~\cite{Ball:2012cx}, and include theory errors (scale and parton density variations) with a central scale fixed to half the hadronic activity in the event ($H_{T}/2$). The transverse momentum for any parton-level jet must satisfy $p_{T} \!>\! 20$~GeV. \label{tab:Xsec}}
\end{table}

\smallskip

\textbf{\textit{Characterization of a boosted four-top system}} -- To demonstrate the feasibility of reconstructing a final-state with four boosted top quarks and its utility in identifying new physics, we leverage significant recent advances in top-tagging~\cite{ATLAS:2015ddu,ATLAS:2022qby,CMS:2017wyc}. 

Our analysis begins by selecting lepton candidates, requiring their transverse momentum $p_T$ to exceed 20~GeV and their pseudo-rapidity to fall within the ranges $|\eta^e| \!<\! 2.47$ and $|\eta^\mu| \!<\! 2.5$ for electrons and muons, respectively. Additional isolation criteria are applied, ensuring that the sum of transverse momenta of all tracks within a cone of radius $\Delta R$ centered on the lepton direction remains below 6\% (4\%) of the electron (muon) transverse momentum, with a cone size determined by $\Delta R = {\rm min}[\Delta R^\mathrm{min},10{\rm \ GeV}/p_T]$ where $\Delta R^\mathrm{min}=0.2$ (0.3). We reconstruct two classes of jets using the anti-$k_T$ algorithm~\cite{Cacciari:2008gp} with radius parameters $R=0.4$ and 1.0, which we refer to as AK4 and AK10 jets, respectively. AK4 jet candidates must satisfy $p_T\!>\!20$ GeV and $|\eta|\!<\!2.5$, while AK10 jets are required to have $p_T \!>\! 350$ GeV, $|\eta|<2.0$, and an invariant mass larger than 40 GeV after applying the soft-drop grooming procedure~\cite{Larkoski:2014wba}. We address overlap between leptons and AK4 jets following the procedure of~\cite{ATLAS:2020hpj}, and AK10 jets within a transverse distance $\Delta R<1$ of a lepton are additionally removed.

We tag AK4 jets as either $b$-jets or non-$b$ jets using a $p_T$-dependent performance similar to the MV2c10 ATLAS algorithm~\cite{ATLAS:2016gsw}. We select a working point that achieves an average tagging efficiency of 77\%, with average mis-tagging rates for $c$-jets, $\tau$-jets and lighter jets of 16.7\%, 4.5\% and 0.75\%, respectively, as determined from top-antitop events at $\sqrt{S}=$ 13~TeV. AK10 jets overlapping with a parton-level top quark within $\Delta R<0.75$ and containing a $B$-hadron are considered to originate from a top quark. Following the performance of the \texttt{HlDNN} and \texttt{ParticleNet} algorithms outlined in~\cite{ATLAS:2024rua} for jets with $p_T>350$~GeV, these AK10 jets are uniformly top-tagged with an efficiency of 80\%. Conversely, AK10 jets not consistent with a top quark are mis-top-tagged with a $p_T$-dependent rate of 10\% (conservative, à la \texttt{HlDNN}) or 5\% (optimistic, à la \texttt{ParticleNet}) on average. 

Our analysis is divided into several signal regions based on the number of top quarks decaying leptonically. In a first category, we select events with \textit{at most} one lepton. This encompasses events with four top quarks decaying hadronically, as well as events with three hadronically-decaying top quarks and one leptonically-decaying top quark, constituting approximately two thirds of all $t\bar{t}t\bar{t}$ events. In a second class of signal regions, we require events to feature two same-sign leptons, aiming to reduce contamination from the $t\bar{t}jj$ background.

In the first category, if an event contains a lepton, it must also feature one additional $b$-tagged AK4 jet that does not overlap with any top-tagged AK10 jets. This ensures the kinematic reconstruction of a leptonically-decaying top quark. The longitudinal momentum of the associated neutrino is determined by assuming that it originates from the decay of an on-shell $W$ boson, which results in two solutions. We choose the one that yields a reconstructed top-quark invariant mass closest to the top pole mass. A first signal region SR1 is dedicated to events featuring three reconstructed top-like objects and one additional (top-tagged or not) AK10 jet. SR1 therefore includes events with either two top-tagged AK10 jets, one additional AK10 jet, and one leptonically-decaying top quark, or events with three top-tagged and one additional AK10 jets without an isolated lepton. Correspondingly, events with four reconstructed top-like objects are placed in a region SR2. It includes events with four top-tagged AK10 jets without a leptonically-decaying top quark, as well as events with three top-tagged AK10 jets and one leptonically-decaying top.

Moreover, we introduce a third signal region, SSL. It is populated by events featuring exactly two leptons with the same electric charge, assumed to originate from two leptonically-decaying top quarks. In addition, we require the presence of at least two AK10 jets and two $b$-tagged AK4 jets that do not overlap with them, along with missing transverse energy larger than 50 GeV. The reconstruction of the two associated neutrinos is performed by minimizing the maximum transverse masses of the two reconstructed $W$ bosons (which corresponds to evaluating the $m_{T2}$ variable for the di-leptonic system~\cite{Lester:1999tx}), and enforcing that these $W$ bosons are on-shell. Among the two possible solutions for the longitudinal component of the neutrinos' momenta, we select the ones with the smallest absolute values. Finally, the reconstruction of the leptonically-decaying top quarks is achieved by pairing the two leading $b$-jets with the leptons and neutrinos, ensuring that the resulting tops have masses closest to the top pole mass.

%%%%%%%%%%%%%%%%%%%%%%%%%%%%%%%%
\begin{figure}
  \centering
  \includegraphics[width=0.49\textwidth]{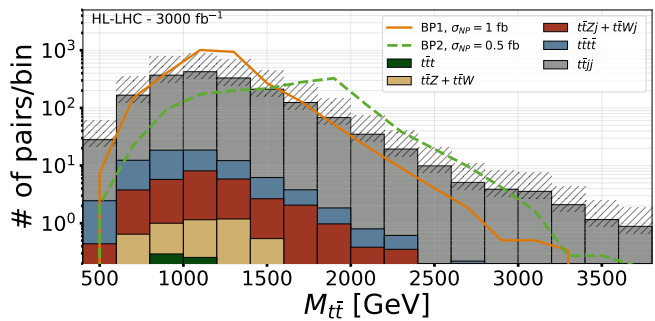}
  \caption{Distribution of the reconstructed top pair invariant mass $M_{t\bar{t}}$, normalized to 3000 fb$^{-1}$, resulting from events passing the SR1 selection with conservative top-tagging performance. Predictions are provided for two signal scenarios (defined in the text) alongside all background contributions accounting for theory uncertainties stemming from scale variation. The last bin incorporates overflow events.\label{fig:signalDistri}}
\end{figure}
%%%%%%%%%%%%%%%%%%%%%%%%%%%%%%%%

At this stage, each selected event falls into at least one of the signal regions and features four top-candidate objects, some of which being top-like (\textit{i.e.}, a reconstructed leptonically-decaying top quark or a top-tagged AK10 jet) and others not (a non-top-tagged AK10 jet). First, we apply an invariant-mass pairing tailored to the production of a pair of top-philic resonances, $pp \!\to\! XX \!\to\! t\bar{t} t\bar{t}$. For each possible pairing among the four top candidates, we compute the two di-top invariant masses and select the combination leading to the smallest difference between these values. If the relative difference satisfies a certain threshold, determined so that 80\% of the signal events from the pair production mechanism meet this criterion, then the two invariant mass values are chosen as proxies for the mass of the $X$ resonance. Otherwise, the event is considered compatible with the associated production mechanism, $pp\to t\bar{t}X \!\to\! t\bar{t} t\bar{t}$. Here, we use as a single proxy for the resonance mass the invariant mass of the top pair with the largest opening angle $\theta$, defined by $\cos\theta = \vec{p}_1 \!\cdot\! \vec{p}_2 / (|\vec{p}_1|\, |\vec{p}_2|)$ where $\vec{p}_{1,2}$ are the three-momenta of the two top candidates. We emphasize that this pair-matching strategy aims only at identifying the relevant invariant mass associated with the $X$ resonance, and is not used in the selection cuts.

\smallskip 

\textbf{\textit{Results and projections}} -- Applying the procedure outlined above, we found that the SR1, SR2 and SSL selection efficiencies are respectively of about 5\%, 1.5\% and 0.5\% for new physics states heavier than 1 TeV, regardless of whether they are produced in pairs or in association with top quarks and after reminding that in the models considered BR$(X\to t\bar{t}) = 1$. Moreover, the efficiency does not strongly depend on the optimistic or conservative scenarios for top-tagging performance, and it sharply decreases for smaller new physics masses. In Fig.~\ref{fig:signalDistri}, we present the distributions of the reconstructed top pair invariant mass $M_{t\bar{t}}$ resulting from the SR1 selection, both for the background and two representative benchmark scenarios featuring a scalar-octet state. For scenario BP1, characterized by $m_{\oct} = 1.3$~TeV and $\yoS=0.25$, pair production dominates, whereas for BP2, with $m_{\oct}=2$~TeV and $\yoS=1$, both pair and associated productions contribute due to the larger coupling value. In the figure, we show the predicted number of signal events after all SR1 selection cuts for the BP1 and BP2 scenarios and an integrated luminosity of 3000~fb$^{-1}$. The signal exhibits a peak at the resonance mass, with a resolution of a few hundred GeV. In contrast, the background, that is primarily driven by $t \bar{t} j j$ production after the mis-tagging of non-top AK10 jets as top candidates, displays a spectrum that smoothly decreases with increasing $M_{t\bar{t}}$ values. Notably, unlike in previous $t \bar t t \bar t$ analyses, contributions from the $ttV$ background are suppressed due to our specific signal reconstruction which relies on four boosted top candidates.

We establish bounds at the $95\%$ confidence level (C.L.) on the signal by leveraging the shape of the $M_{t\bar{t}}$ distribution, employed as input to the \pyhf\ software~\cite{Heinrich:2021gyp} for limit setting. Given the substantial theoretical uncertainties on the dominant $t\bar{t}jj$ background contribution, we incorporate a correlated scale factor on the background predictions.

\begin{table}\renewcommand{\arraystretch}{1.25}\setlength{\tabcolsep}{6pt}
\resizebox{0.99\columnwidth}{!}{%
  \begin{tabular}{c|cccc|cccc}
     & \multicolumn{4}{c|}{BP1} & \multicolumn{4}{c}{BP2} \\
     Top-tag. & \multicolumn{2}{c}{Optimistic} & \multicolumn{2}{c|}{Conservative} & \multicolumn{2}{c}{Optimistic} & \multicolumn{2}{c}{Conservative} \\
     $\mathcal{L}$ [fb$^{-1}$] & 400 & 3000 & 400 & 3000 & 400 & 3000 & 400 & 3000 \\   
    \toprule
    SR1 & 1.35 & 0.52 & 1.69 & 0.64 & 0.68 & 0.24 & 0.82 & 0.30 \\
    SR2 & 0.64 & 0.26 & 0.75 & 0.36 & 0.51 & 0.14 & 0.61 & 0.20 \\
    SSL & 0.97 & 0.27 & 0.97 & 0.27 & 1.13 & 0.29 & 1.12 & 0.28 \\
  \end{tabular}}
  \caption{Upper limits on the new physics cross section (in fb), considering the benchmark scenarios BP1 and BP2, derived from our analysis strategy and its three signal regions. The scenarios lead to a signal dominated by pair production of a new scalar (BP1), and made of an admixture of its pair and associated production (BP2). Results are shown for both optimistic and conservative top-tagging performance, with luminosities of 400 and 3000 fb$^{-1}$. \label{tab:CSlim}}
\end{table}

%%%%%%%%%%%%%%%%%%%%%%%%%%%%%%%%
\begin{figure*}
  \centering
  \includegraphics[width=0.48\textwidth]{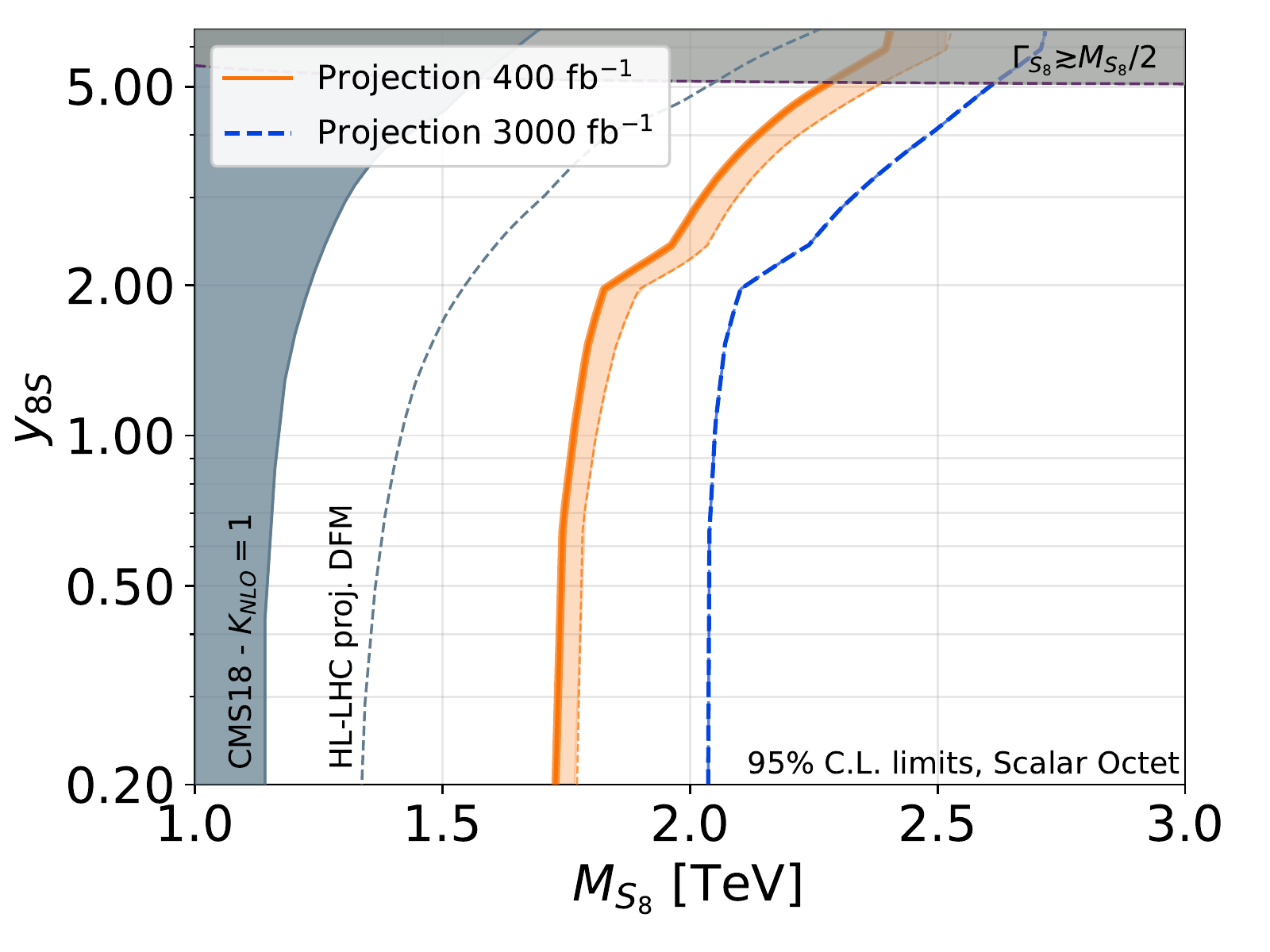}\hfill
  \includegraphics[width=0.48\textwidth]{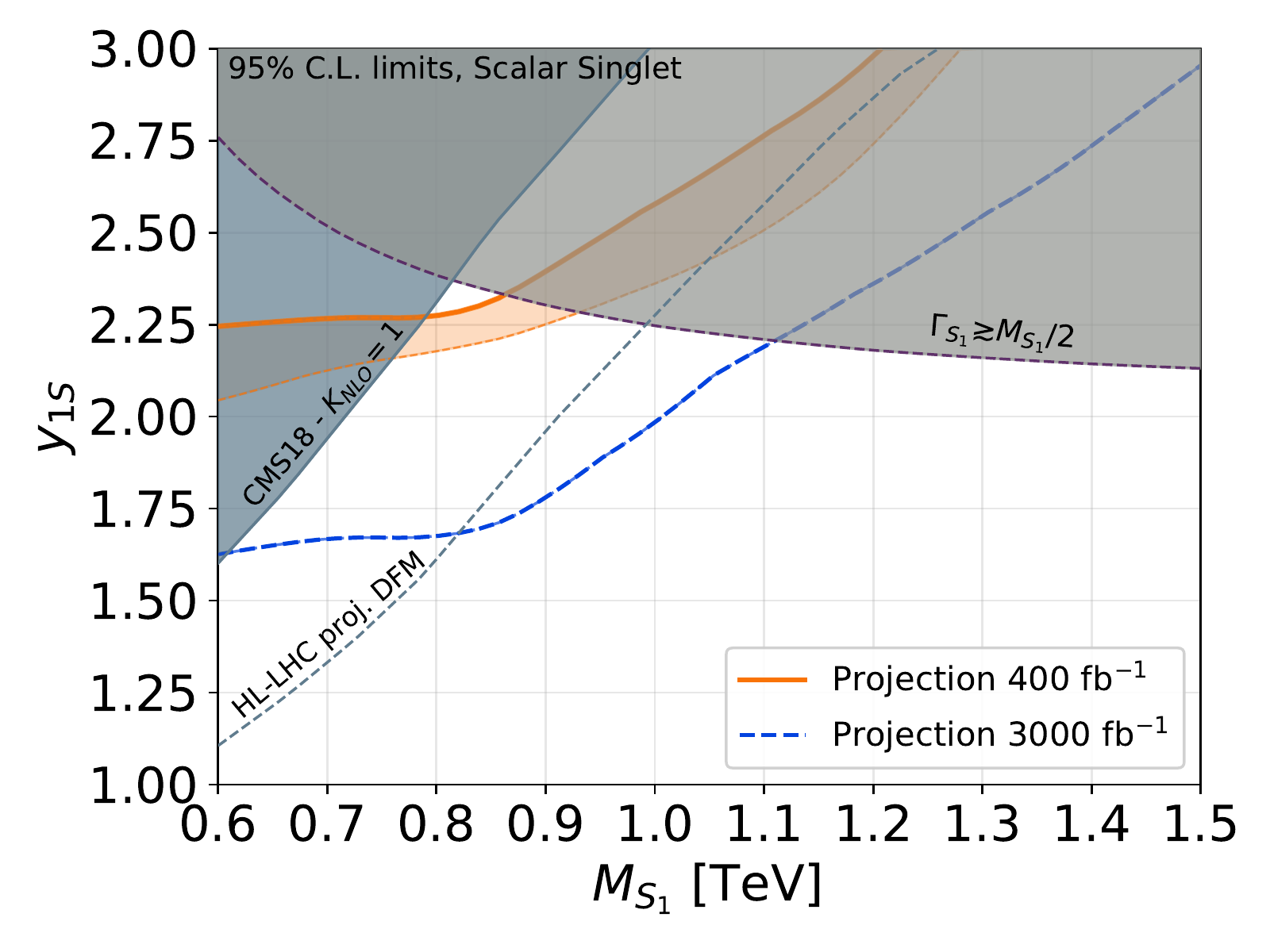}
  \caption{Projected 95\% C.L. exclusions for the scalar color-octet (left) and scalar color-singlet (right) simplified models, presented in mass {\it versus} coupling planes, considering luminosities of $400$ (orange) and $3000$~fb$^{-1}$ (dashed blue). The solid orange lines represent exclusions obtained with conservative top-tagging performance, while the dashed orange lines refer to a more optimistic assumption. These limits are compared to the currently excluded region of the parameter space, obtained from a recast of the CMS-TOP-18-003 analysis (dark shades) \cite{Darme:2021gtt}, and its naive extrapolation for the HL-LHC with 3000~fb$^{-1}$ (dashed gray). In addition, light gray regions in the top-right corners of the figures represent configurations in which the width of the new scalar is very large, rendering our approach unreliable.\label{fig:ylimOctetSinglet}}
\end{figure*}
%%%%%%%%%%%%%%%%%%%%%%%%%%%%%%%%

Our analysis commences with an examination of the two benchmark scenarios BP1 and BP2, our findings being summarized in Table~\ref{tab:CSlim}. Results are presented in terms of projected sensitivity for both the LHC run~3 (with a luminosity of 400~fb$^{-1}$) and its HL-LHC upgrade (with a luminosity of 3000~fb$^{-1}$). Sensitivity is expressed as an upper limit on the total signal production cross section (times branching ratios into top quarks). Remarkably, irrespective of the assumed top-tagging performance, we observe a dramatic enhancement of almost two orders of magnitude compared to projections from typical analyses currently conducted at the LHC~\cite{CMS:2017ocm, ATLAS:2018alq, CMS:2019rvj, ATLAS:2023taw}. This originates from the fact that these analyses rely solely on the multiplicity of the reconstructed final-state objects (including potentially fat jets), and additionally account for the invariant mass of a pair of fat jets in the case of single production of a top-philic resonance. In contrast, we have designed a method sensitive to both single and pair production of a top-philic resonance, showing for the first time that a full exploitation of the kinematics reconstruction of the four-top final state was possible. To illustrate this, a straightforward extrapolation of the CMS-TOP-18-003 analysis~\cite{CMS:2019rvj} in which signal and background events and their errors are rescaled linearly to a luminosity of 3000~fb$^{-1}$ would constrain new physics contributions to the four-top production rate to approximately 10~fb~\cite{Darme:2021gtt}. In contrast, our proposed search strategy yields bounds that are 30 to 60 times stronger, ranging from 0.15 to 0.3 fb, with the exact improvement contingent on the assumptions on top-tagging performance. For models in which single production dominates, a further comparison can be made with the ATLAS-EXOT-2022-14 analysis~\cite{ATLAS:2023taw} that probes singly-produced top-philic resonances leading to a four-top signal, utilizing a pair of fat jets and the corresponding invariant mass (\textit{i.e.}\ without top-tagging). While an interpretation directly applicable to the models studied in this work is not possible, we use the top-philic $Z'$ model examined in~\cite{ATLAS:2023taw} to assess that our technique still allows for a factor of 3 improvement, which could approximately be applied to the $S_1$ model. Further details will be provided in an upcoming publication. This result therefore not only underscores the feasibility of conducting a search for new physics in events with four boosted top quarks, but it also emphasizes its unparalleled efficacy in exploiting the LHC's potential for probing physics beyond the SM.

To showcase the effectiveness of our approach, we examine the sensitivity in Fig.~\ref{fig:ylimOctetSinglet} to the color-octet (left) and color-singlet (right) simplified models under consideration. We delineate its projected reach for the LHC run-3 (orange bands encompassing different choices for top-tagging performance), and for the HL-LHC and optimistic top tagging performance (dashed blue curves). These exclusions are juxtaposed with current limits derived from the recast of the results of LHC run 2's CMS-TOP-18-003 analysis (dark shaded regions on the left of the plots, sourced from~\cite{Darme:2020hxc, Darme:2021gtt}), and their extrapolation for the HL-LHC luminosity (dashed gray curves, from \cite{Araz:2019otb, Darme:2020hxc, Darme:2021gtt}). Our strategy remarkably demonstrates a substantial enhancement in mass reach compared to traditional searches for new physics with four-top production. For instance, we have found that, from LO rates, heavy color-octet scalars could potentially be excluded for masses up to 2.1~TeV, even with moderately small Yukawa couplings $\yoS$. This consists of a significant improvement over current search strategies employed by the ATLAS and CMS collaborations, which currently reach only masses up to approximately 1.2~TeV. This corresponds to a region of the parameter space where new physics contributions to four-top production are induced by scalar-octet pair production. 

Currently, color-singlet states face mass limits ranging from 600 to 800~GeV, given Yukawa couplings falling in the 1.5--2.5 regime. Extrapolating these limits to the HL-LHC luminosity extends them to 600--1000~GeV, possibly with slightly lower couplings. Consequently, new physics four-top events may not necessarily be highly boosted, posing a challenge for our analysis. Our findings confirm this limitation, although it additionally largely stems from large theory uncertainties on the background. While anticipated improvements in sensitivity by applying our approach to the LHC run~3 only marginally increase the reach for color-singlets with masses of 800--900~GeV, it yields exclusion limits substantially complementary to those obtained from traditional searches at the HL-LHC, particularly for singlet masses above 1~TeV. Although not as dramatic as in the case of scalar-octet states, our approach therefore still holds promise for improving searches for top-philic color-singlets with masses exceeding 1~TeV at the HL-LHC.

Furthermore, our findings exhibit minimal sensitivity to future top-tagging performance once cast in a mass \textit{versus} coupling plane, by virtue of the dependence of the fiducial signal cross section on the new physics parameters and that we expect only $\mathcal{O} (1)$ background events with the most optimistic top-tagging performance. This is exemplified at a luminosity of 400 fb$^{-1}$, where the reaches under optimistic and conservative top-tagging assumptions are delineated by the two edges of the orange band. With this band being relatively narrow, it is evident that the influence of top-tagging performance is modest. However, achieving sensitivity beyond the 2~TeV barrier at the HL-LHC may pose a challenge and require further improvements in top-tagging efficiency. 

\smallskip

\textbf{\textit{Conclusion}} -- In this letter, we introduced a pioneering analysis strategy designed to explore new physics through four-top events at the LHC. Leveraging significant progress in top-tagging techniques, our method enables the efficient reconstruction of boosted four-top systems. Subsequently, it allows for the characterization of hypothetical top-philic particles by effectively controlling the associated SM background.

To demonstrate the effectiveness of our approach, we applied it to two scenarios with new top-philic scalars. Our results revealed a significant enhancement in sensitivity to these scalars at the LHC, with bounds on associated cross sections improving by a notable factor of a few for color singlets up to a factor of 60 for color octets compared to existing searches. This brings the bounds on these rates below the fb level, providing unprecedented insights into new top-philic resonances. Specifically, our analysis indicates that color-charged scalars could potentially be constrained beyond the 2~TeV barrier, while color-singlet scalars with substantial top Yukawa couplings may be restricted to the TeV regime.

Importantly, we emphasize that future improvements in top-tagging algorithms have the potential to further strengthen these limits, highlighting the promise of our approach in pushing the boundaries of new physics exploration at the LHC.

\newpage

%%%%%%%%%%%%%%%%%%%%%%%%%%%%%%%%%%%%%%%%%%%%%%%%%%%%%%%%%%%%%%%%%%%%%%%%%%%%%%%%
\textit{\textbf{Acknowledgments}} -- We thank C.~Degrande and S.~Vatani for valuable contributions during the early stage of this work, and J.Y.~Araz for technical support with \ma. Computational resources have been provided by the Consortium des Équipements de Calcul Intensif (CÉCI), funded by the Fonds de la Recherche Scientifique de Belgique (F.R.S.-FNRS) under Grant 2.5020.11 and by the Walloon Region. MM is a Research Fellow of the F.R.S.-FNRS, through the grant \textit{Aspirant}; LD has been supported by the European Union's Horizon 2020 research and innovation programme under the Marie Skłodowska-Curie grant agreement 101028626 from 01.09.2021 to 31.08.2023; BF has been supported by Grant ANR-21-CE31-0013 from the \emph{Agence Nationale de la Recherche} (ANR), France; HL is supported by an F.S.R.~postdoctoral fellowship of the Université Catholique de Louvain. The work of HL, MM, JT has also been supported by the 4.4517.08 IISN-F.N.R.S convention.

\bibliographystyle{JHEP}
\bibliography{bibliography.bib}

\end{document}